# Engineering cavity-field states by projection synthesis

R. M. Serra, N. G. de Almeida, C. J. Villas-Bôas, and M. H. Y. Moussa

*Departamento de Física, Universidade Federal de São Carlos, Via Washington Luis, km 235, São Carlos 13565-905, SP, Brazil*

(Received 10 December 1999; published 13 September 2000)

We propose a reliable scheme for engineering a general cavity-field state. This is different from recently presented strategies, where the cavity is supposed to be initially empty and the field is built up photon by photon through resonant atom-field interactions. Here, a coherent state is previously injected into the cavity. So, the Wigner distribution function of the desired state is constructed from that of the initially coherent state. Such an engineering process is achieved through an adaptation of the recently proposed technique of projection synthesis to cavity QED phenomena.

The program of quantum state engineering has recently become increasingly important for pursuing some striking proposals in nowadays theoretical physics. Such proposals work on the possibility of nonlocality, a crucial character exhibited by an entangled state, and this seems to open the way for a new technology based on microscopic physics. The advent of quantum computation [1], quantum communication [2], and teleportation [3] brought together the necessity of tailoring specific quantum states which are envisaged to play the role of either a qubit or a nonlocal quantum channel. Moreover, the recently proposed technique of *projection synthesis* [4] requires the engineering of some exotic quantum states for measuring particular properties of the radiation field, such as its phase [4] or its $Q$ function [5]. The projection synthesis technique has also been considered itself for engineering [6] and teleporting [7] a running wave superposition state.

Schemes for engineering a superposition state of the radiation field has been presented in the realm of cavity QED phenomena [8]. Vogel *et al.* [9] have basically employed a resonant atom-field interaction for build up an arbitrary trapped field in an empty cavity. An alternative proposal [10], based on Ref. [9], has been presented considering both resonant and dispersive atom-field interaction for preparing a general cavity-field superposition state. In Refs. [9,10] a Ramsey zone, placed in the way of the atoms to the cavity, is supposed to prepare the particular atomic superposition states which are required to properly build up the cavity field, photon by photon.

In the present work we have transposed the projection synthesis technique from its original running wave domain to cavity QED phenomena. This is different from the technique used in [9]; a coherent state is previously injected into the cavity through a wave guide which connects it to a monochromatic source $S$ (see Fig. 1). Thus, the Wigner distribution function of the desired state is sculptured, through the projection synthesis technique, from that representing the initial coherent state. The atoms work as quantum chisels on the Wigner distribution for the initial coherent state as shown in the experimental setup depicted in Fig. 1. Each two-level atom (excited $|e\rangle$ and ground $|g\rangle$ states) is initially prepared in a coherent superposition, in Ramsey zone $R_1$, before crossing cavity $C$. A stream of $M$ atoms is thus made to interact resonantly with the coherent state injected into the cavity. After interacting with the cavity field the atoms are made to cross an additional Ramsey zone $R_2$ in their way to the detection chambers $D$ ($D_e$ and $D_g$, for ionizing the states $|e\rangle$ and $|g\rangle$, respectively). Each atom is supposed to be detected in a particular superposition state before the subsequent atom enters $C$. The detection of these atomic superposition states is envisaged to synthesize the projection of the leaving cavity field in the desired sculptured state, through Ramsey zone $R_2$ and detectors $D$, as shown below.

Considering the $k$th step of this process, let us describe the state of the cavity field, after the injection and detection of the $(k-1)$th atom, as

$$|\Psi_f^{(k-1)}\rangle = \sum_{n=0}^{\infty} \Lambda_n^{(k-1)}|n\rangle, \qquad (1)$$

with $|n\rangle$ representing an $n$-photon Fock state of the cavity mode. The subscript $f$ indicates the field state whereas the superscript $k-1$ indicates the $(k-1)$th atom. After the interaction of the $k$th atom [initially prepared by $R_1$ in the state $\mathcal{N}_{\beta_k}(|e\rangle_k + \beta_k|g\rangle_k)$, where $\mathcal{N}_{\beta_k} = (1+|\beta_k|^2)^{-1/2}$ and $\beta_k$ is a complex number] with the field (1), the atom-field system $(a,f)$ evolves to the entangled state

$$|\psi_{a,f}^{(k)}\rangle = \mathcal{N}_{\beta_k} \sum_{n=0}^{\infty} \Lambda_n^{(k-1)}(C_n^{(k)}|e\rangle_k|n\rangle - iS_n^{(k)}|g\rangle_k|n+1\rangle$$
$$+ \beta_k C_{n-1}^{(k)}|g\rangle_k|n\rangle - i\beta_k S_{n-1}^{(k)}|e\rangle_k|n-1\rangle), \qquad (2)$$

where $C_n^{(k)} = \cos(\sqrt{n+1}\,\Omega\tau_k)$ and $S_n^{(k)} = \sin(\sqrt{n+1}\,\Omega\tau_k)$, $\Omega$ being the atom-cavity-mode coupling strength, and $\tau_k$ the interaction time of the $k$th atom with the cavity field.

Now, following the steps which lead us to describe the field state (1) after detecting the $(k-1)$th atom, we have to synthesize the projection of the atom-field entangled state (2) into a particular atomic superposition state of the $k$th atom $|\phi_a^{(k)}\rangle = \mathcal{N}_{\varepsilon_k}(|e\rangle_k + \varepsilon_k^*|g\rangle_k)$ with $\mathcal{N}_{\varepsilon_k} = (1+|\varepsilon_k|^2)^{-1/2}$. In this way, the field state after the detection of the $k$th atom reads

$$|\Psi_f^{(k)}\rangle = \mathcal{N}_k \langle \phi_a^{(k)}|\psi_{a,f}^{(k)}\rangle = \sum_{n=0}^{\infty} \Lambda_n^{(k)}|n\rangle. \quad (3)$$

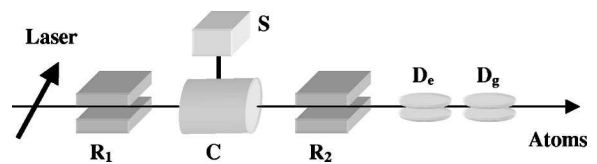

FIG. 1. Sketch of the experimental setup for sculpturing a cavity-field state by projection synthesis.

The coefficients $\Lambda_n^{(k)}$ result from those in Eq. (2) through the recurrence formula $\Lambda_n^{(k)} = \mathcal{N}_k \Gamma_n^{(k)}$, where

$$\Gamma_n^{(k)} = (C_n^{(k)} + \varepsilon_k \beta_k C_{n-1}^{(k)})\Lambda_n^{(k-1)} - i\beta_k S_n^{(k)} \Lambda_{n+1}^{(k-1)}$$
$$- i\varepsilon_k(1-\delta_{n,0})S_{n-1}^{(k)}\Lambda_{n-1}^{(k-1)}, \quad (4)$$

with $\Lambda_n^{(0)} = \exp(-|\alpha|^2/2)\alpha^n/\sqrt{n!}$. The normalization constant reads

$$\mathcal{N}_k = \left[\sum_{n=0}^{\infty}|\Gamma_n^{(k)}|^2\right]^{-1/2}. \quad (5)$$

The complex number $\varepsilon_k$ results from the rotation of the $k$th atom is the Ramsey zone $R_2$ in order to synthesize the projection of the entangled state $|\psi_{a,f}^{(k)}\rangle$ as in Eq. (3). The Ramsey zone $R_2$ is needed in order to adjust the measurement of this special superposition, playing the role of atomic states ''polaryzers'' and ''analyzers'' which, combined with detectors $D$, allow us to analyze an arbitrary superposition of $|e\rangle$ and $|g\rangle$. As considered by Freyberger [11], such a measurement works as follows: The Ramsey zone $R_2$ is appropriately adjusted in a way that a two-level atom which crosses it in the superposition $|\phi_a^{(k)}\rangle$ undergoes a unitary transformation to the state $|e\rangle_k$. As a consequence, the $k$th atom leaves the Ramsey zone in the state $|g\rangle_k$ if it crosses it in the state orthogonal to $|\phi_a^{(k)}\rangle$. After crossing $R_2$, the $k$th atom is thus counted with high efficiency by the ionization detection chambers $D$. However, the $k$th atom will be in a superposition of these orthogonal states, so that in some cases the detector will deliver a click (measuring $|e\rangle_k$) and in some it may not click (measuring $|g\rangle_k$). Once we register a click, the $k$th atom has been projected in the required superposition $|\phi_a^{(k)}\rangle$. So, the analyzer is able to measure observables of the form $|\phi_a^{(k)}\rangle\langle \phi_a^{(k)}|$, which represent projection operators with measurable eigenvalues 1 (click) and 0 (no click). The coefficient $\varepsilon_k$ is left to be determined through those ($d_n$) of the desired state $|\psi_d\rangle = \sum_{n=0}^{N_d} d_n|n\rangle$.

Since the projection of the $M$th atom in a particular state is supposed to finish the sculpture process, the following equality must be satisfied:

$$\mathcal{N}_M \langle \phi_a^{(M)}|\psi_{a,f}^{(M)}\rangle = \sum_{n=0}^{\infty} \Lambda_n^{(M)}|n\rangle = \sum_{n=0}^{N_d} d_n|n\rangle, \quad (6)$$

requiring that $\Lambda_n^{(M)} \approx 0$ when $n \geq N_d + 1$. Such approximation brings about a nonunity *fidelity* [12] for the sculptured state, which is defined by the overlap between the desired and the sculptured state (the field state after detecting the $M$th atom), following the expression:

$$\mathcal{F} = |\langle \psi_d|\psi_f^{(M)}\rangle|^2 = \frac{|\sum_{n=0}^{N_d} d_n^* \Gamma_n^{(M)}|^2}{\sum_{l=0}^{\infty}|\Gamma_l^{(M)}|^2}. \quad (7)$$

The total probability for successfully sculpturing the desired state reads $\mathcal{P} = \Pi_{k=1}^M P_k$, where $P_k$ accounts for the probability to synthesize the $k$th-atom-field entanglement into a particular atomic superposition, $|\langle \phi_a^{(k)}|\psi_{af}^{(k)}\rangle|^2$, following the expressions (4) and (5):

$$P_k = \mathcal{N}_{\varepsilon_k}^2 \mathcal{N}_{\beta_k}^2 \sum_{n=0}^{\infty} |\Gamma_n^{(k)}|^2. \quad (8)$$

For an appropriated choice of the average excitation of the coherent state, $|\alpha|^2 = \bar{n}_\alpha$, we see from the recurrence formula (4) and the definition of the coefficients $\Lambda_n^{(0)}$, that the coefficients $\Gamma_n^{(M)}$ depend on the powers of $\alpha$, as $\alpha^\xi/\sqrt{\xi!}$, with $n-M \leq \xi \leq n+M$. In fact, it is straightforward to conclude that, after one application of formula (4) the coefficients $\Gamma_n^{(M)}$ are proportional to those $\{\Lambda_{n-1}^{(M-1)}, \Lambda_{n-1}^{(M-1)}, \Lambda_n^{(M-1)}, \Lambda_{n+1}^{(M-1)}\}$; after two applications of formula (4) it follows that $\Gamma_n^{(M)} \propto \{\Lambda_{n-2}^{(M-2)}, \Lambda_{n-1}^{(M-2)}, \Lambda_n^{(M-2)}, \Lambda_{n+1}^{(M-2)}, \Lambda_{n+2}^{(M-2)}\}$ and after $M$ applications of such formula we finally obtain $\Gamma_n^{(M)} \propto \{\Lambda_{n-M}^{(0)}, \Lambda_{n-M+1}^{(0)}, \ldots, \Lambda_n^{(0)}, \ldots, \Lambda_{n+M-1}^{(0)}, \Lambda_{n+M}^{(0)}\}$. Therefore, from the definition of $\Lambda_n^{(0)}$, we note that $\Gamma_n^{(M)}$ depends on the powers of $\alpha$ as mentioned above. So, the choice of the average excitation $\bar{n}_\alpha$ in a way that $P(N_d - M + 1) = |\langle N_d - M + 1|\alpha\rangle|^2 \approx 0$ (satisfying the requirement that $\Lambda_n^{(M)} \approx 0$ when $n \geq N_d + 1$), results in a higher fidelity $\mathcal{F}$ at the expenses of a lower probability $\mathcal{P}$. In fact, it is evident from the denominator of Eq. (7) that the lesser the number $l$ in the sum of significant coefficients, the greater the fidelity. On the other hand, from Eq. (8) we observe that the probability $\mathcal{P}$ is directly proportional to the powers of $\alpha$. As a consequence, Eqs. (7) and (8) impose a fidelity-probability rate, $\mathcal{R} \equiv \mathcal{F}\mathcal{P}$, a cost-benefit estimative for sculpturing the desired state. In order to maximize the rate $\mathcal{R}$, we have to play with all the variables appearing in the recurrence formula (4): the average excitation $\bar{n}_\alpha$ of the initial coherent state injected into the cavity, the atom-field interaction times $\tau_k$, and the Ramsey zones parameters $\varepsilon_k$ and $\beta_k$. Beginning with $\bar{n}_\alpha$, we note that a good strategy to maximize $\mathcal{R}$ consists in starting with the choice of $\bar{n}_\alpha$ so that $P(N_d - M + 1) \approx 0$, and then proceed to maximize the rate $\mathcal{R}$, increasing $\bar{n}_\alpha$ at the expenses of the fidelity. Concerning the interaction times $\tau_k$, in the scheme of Vogel *et al.* these variables are either fixed *a priori* or appropriately chosen in order to maximize the probability for achieving the desired state. Here, however, when the variables $\tau_k$ are not fixed *a priori* (a procedure depending on the experimental capabilities), they are chosen in order to maximize the rate $\mathcal{R}$. Finally, accounting for the choice of the Ramsey zones parameters $\varepsilon_k$ and $\beta_k$, they follow from a particular solution of the equality

(6) from which, on the requirement $\Lambda_n^{(M)} \approx 0$ when $n \geq N_d + 1$, results the set of $N_d + 1$ equations

$$d_{N_d} = \mathcal{N}_M [(C_{N_d}^{(M)} + \varepsilon_M \beta_M C_{N_d-1}^{(M)}) \Lambda_{N_d}^{(M-1)} - i\beta_M S_{N_d}^{(M)} \Lambda_{N_d+1}^{(M-1)} - i\varepsilon_M S_{N_d-1}^{(M)} \Lambda_{N_d-1}^{(M-1)}],$$

$$\vdots = \vdots$$

$$d_n = \mathcal{N}_M [(C_n^{(M)} + \varepsilon_M \beta_M C_{n-1}^{(M)}) \Lambda_n^{(M-1)} - i\beta_M S_n^{(M)} \Lambda_{n+1}^{(M-1)} - i\varepsilon_M S_{n-1}^{(M)} \Lambda_{n-1}^{(M-1)}], \quad (9)$$

$$\vdots = \vdots$$

$$d_0 = \mathcal{N}_M [(C_0^{(M)} + \varepsilon_M \beta_M) \Lambda_0^{(M-1)} - i\beta_M S_0^{(M)} \Lambda_1^{(M-1)}].$$

In Ref. [9] a similar set of equations is obtained, but with $M = N_d$, since this is the number of atoms required to build up the desired field state from the vacuum. In that case, the set of linear equations, similar to Eq. (9), results in an algebraic equation of degree $N_d$ in the variables $\varepsilon_k$. In the present case, to solve the set of equations (9) we apply the recurrence formula (4) $M - 1$ times in order to express the unknown coefficients $\Lambda_n^{(k)}$ in terms of the known values of the coefficients of the coherent state, $\Lambda_n^{(0)}$. In this way we obtain a nonlinear system whose free parameters are $\varepsilon_1, \ldots, \varepsilon_M$ and $\beta_1, \ldots, \beta_M$, which indicate the particular rotation each atom must undergo in both Ramsey zones. Such variables are obtained from the known coefficients $d_n$ and $\Lambda_n^{(0)}$. The solubility of a nonlinear system can be guaranteed if the number of equations is equal to the number of variables, the later being the parameters of both Ramsey zones. One of the equations in system (9) has to be used to obtain the normalization constant $\Pi_{k=1}^M \mathcal{N}_k$ and each atom carries two free parameters ($\varepsilon_k, \beta_k$). Therefore, the minimum number of atoms necessary to guarantee the solution of system (9) must be $M = \text{int}[(N_d + 1)/2]$. This conclusion is due to the fact that in our scheme we start from a coherent field state previously injected into the cavity, differently from [9] the scheme of Vogel *et al.* which starts with the vacuum state requiring $N_d$ atoms. So, the advantage of the present technique is that it requires just half of the number of atoms needed in [9] for sculpturing a desired state and, consequently, half of the experimental running time required for realizing the process.

Despite the fact that we are considering here the ideal case, it is worth mentioning that two major error sources are present when treating the real situation: the efficiency of the atomic detection chambers and the running time of the experiment. The latter is due to the relaxation times associated with the quantum systems since dissipation mechanisms are induced by their inevitable coupling to the environment. So, when accounting for these error sources our scheme results in a higher fidelity (now due to the errors introduced by the environment [12]) of the engineered state, compared with that in [9] which requires twice the number of atoms $M$ and, consequently, twice the running time of the experiment. It is worth noting that the additional Ramsey zone required in the present scheme (for the projection synthesis) and absent from the setup in [9], does not comprehend a significant error source. Since the position of each atom can be determined at any time between preparation and detection with a precision better then 1 mm, it allows us to fire microwave pulses in both Ramsey zones exactly when the atoms reach the corresponding position (with the possibility of exposing successive atoms to different interactions) [13]. So, the errors introduced in Ramsey zones are significantly smaller than those coming from the atomic decay and the detection chambers (detection efficiency of about 35% [13]).

To illustrate the present technique we now proceed to sculpture the truncated phase state ($N_d = 4$)

$$|\psi_d\rangle = \frac{1}{\sqrt{5}} \sum_{n=0}^{4} |n\rangle, \quad (10)$$

which requires just $M = 2$ two-level atoms, instead of $M = 4$ as in [9]. As mentioned above, we start with the average excitation $\bar{n}_\alpha$ leading to the higher fidelity: requiring that $P(3) = |\langle 3|\alpha\rangle|^2 \approx 0$, it follows the average excitation $\bar{n}_\alpha = 0.36$. For each value of $\bar{n}_\alpha$ (choosing $\alpha$ as a real parameter) we proceed to calculate the interaction parameters $\Omega \tau_1$ and $\Omega \tau_2$ which maximize the rate $\mathcal{R}$. In Table I we show the rate $\mathcal{R}$ associated with each value of $\bar{n}_\alpha$, from that which maximize the fidelity, 0.36, to that showing a continuous decreasing of the rate, 4.0. In addition, a set of numerically obtained roots $(\varepsilon_1, \beta_1; \varepsilon_2, \beta_2)$ of the equations (9), results when fixing a given couple of interaction parameters $(\Omega\tau_1, \Omega\tau_2)$, and we have to choose the one which maximizes $\mathcal{R}$.

TABLE I. The probability $\mathcal{P}$, fidelity $\mathcal{F}$, and rate $\mathcal{R} = \mathcal{PF}$, for each value of averaged excitation number $\bar{n}_\alpha$. The interaction parameters $\Omega\tau_1$ and $\Omega\tau_2$ in this table correspond to values which maximize the rate $\mathcal{R}$ for each $\bar{n}_\alpha$.

| $\bar{n}_\alpha$ | $\Omega\tau_1$ | $\Omega\tau_2$ | $\mathcal{P}$ | $\mathcal{F}$ | $\mathcal{R}$ |
|---|---|---|---|---|---|
| 0.3600 | 5.9000 | 5.1000 | 0.0468 | 0.9826 | 0.0460 |
| 0.6400 | 3.1000 | 4.1000 | 0.2148 | 0.9983 | 0.2144 |
| 1.0000 | 3.2000 | 4.1000 | 0.2910 | 0.9906 | 0.2883 |
| 1.4400 | 4.1000 | 4.4000 | 0.2319 | 0.9827 | 0.2279 |
| 1.9600 | 5.7000 | 4.3000 | 0.4364 | 0.9770 | 0.4264 |
| 2.5600 | 5.8000 | 4.2000 | 0.5590 | 0.9459 | 0.5288 |
| 3.2400 | 4.9000 | 4.2000 | 0.4295 | 0.9187 | 0.3946 |
| 4.0000 | 5.1000 | 4.2000 | 0.2028 | 0.7953 | 0.1613 |

TABLE II. The Ramsey zones parameters [$\varepsilon_k = |\varepsilon_k| \exp(i\theta_k)$, $\beta_k = |\beta_k| \exp(i\varphi_k)$] and interaction times ($\tau_k$) needed to obtain the truncated phase state for an initial coherent state associated with $\bar{n}_\alpha = 2.56$.

| $k$ | $\Omega\tau_k$ | $|\varepsilon_k|$ | $\theta_k$ | $|\beta_k|$ | $\varphi_k$ | $P_k$ |
|---|---|---|---|---|---|---|
| 1 | 5.800 | 0.4247 | 4.7124 | 0.7684 | 4.7124 | 0.7576 |
| 2 | 4.200 | 0.4616 | 4.7124 | 0.6583 | -1.5708 | 0.7379 |

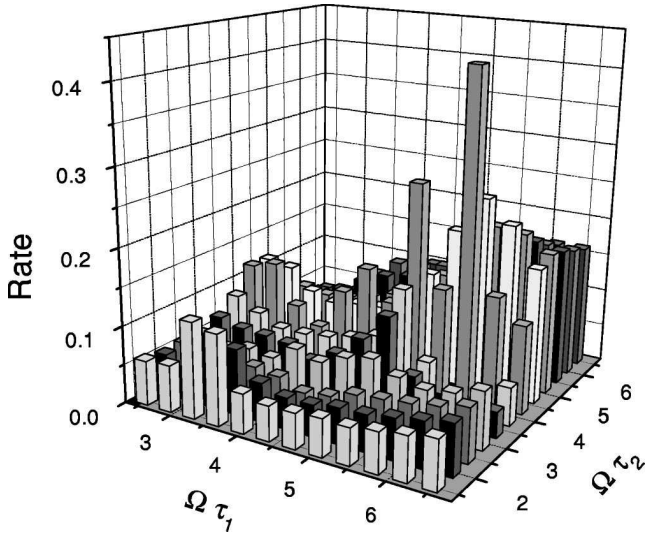

FIG. 2. Histogram of the rate $\mathcal{R} = \mathcal{PF}$, in the $\Omega\tau_1 \times \Omega\tau_2$ plane, for $\bar{n}_\alpha = 2.56$.

We note that for engineering the state (10) through the scheme of Vogel *et al.* results in the probability 0.2833 for the unity fidelity. From Table I, it follows from $\bar{n}_\alpha = 1$ that when $\mathcal{P} = 0.2910$, approximately the probability obtained through the scheme of Vogel *et al.*, the resulting fidelity $\mathcal{F} = 0.9906$, approaches unity. However the fidelity-probability rate for these values reads $\mathcal{R} = 0.2883$, a result which can be considerably increased when choosing $\bar{n}_\alpha = 2.56$. This latter choice results in a higher rate $\mathcal{R} = 0.5288$, which follows from a remarkable probability for successfully sculpturing the desired state, $\mathcal{P} = 0.5590$, and still a considerable fidelity, $\mathcal{F} = 0.9459$. In Table II we display the roots $(\varepsilon_1, \beta_1; \varepsilon_2, \beta_2)$, associated with $\bar{n}_\alpha = 2.56$, which leads to the maximal rate $\mathcal{R} = 0.5288$.

In Fig. 2 we display the histogram of the rate $\mathcal{R}$, in the $\Omega\tau_1 \times \Omega\tau_2$ plane. The higher peak in $\mathcal{R}(\Omega\tau_1, \Omega\tau_2)$ corresponds to the values of $\Omega\tau_1$ and $\Omega\tau_2$ shown in Table II. This figure shows the process of the maximization of $\mathcal{R}$ for a given average excitation number. Finally, in Figs. 3(a)–3(c) we display the sculpture process of the desired truncated phase state (10) from the Wigner distribution function of the initial coherent state (associated to $\bar{n}_\alpha = 2.56$) given by the Gaussian in Fig. 3(a). This Gaussian is shifted from the position origin as $W(p,q) = (2/\pi)\exp[-(q+\alpha)^2 - p^2]$. Figures 3(b) and 3(c) display the sculpture process after the projection synthesis of the first and second atoms, respectively. In Fig. 3(d) the desired state in (10) is exactly displayed, which differs from the sculptured state due to its nonunity fidelity. As we have stressed above it is possible through the present scheme to sculpture the desired state with higher fidelity; however, at the expense of a smaller probability for successfully achieving the process. So, the sculpture technique depends on a cost-benefit estimative, here defined as the fidelity-probability rate, which is up to the sculptor necessi-ties. An interesting point for future investigation consists in account for the noise effects on the sculpture process, i.e., the influence of the errors arising from the inevitable coupling of the required quantum systems to the environment and also the detector inefficiency. When treating the noise effects on the sculpture process, the fidelity defined in Eq. (7) remains exactly the same, but the sculptured field state, after the detection of the $M$th atom, will be entangled to the environmental states. So, to the best of our knowledge the ''sculptured'' cavity field will be represented by a statistical mixture $\rho_f$, whereby the fidelity turns out to be $\mathcal{F} = \langle \psi_d | \rho_f | \psi_d \rangle$.

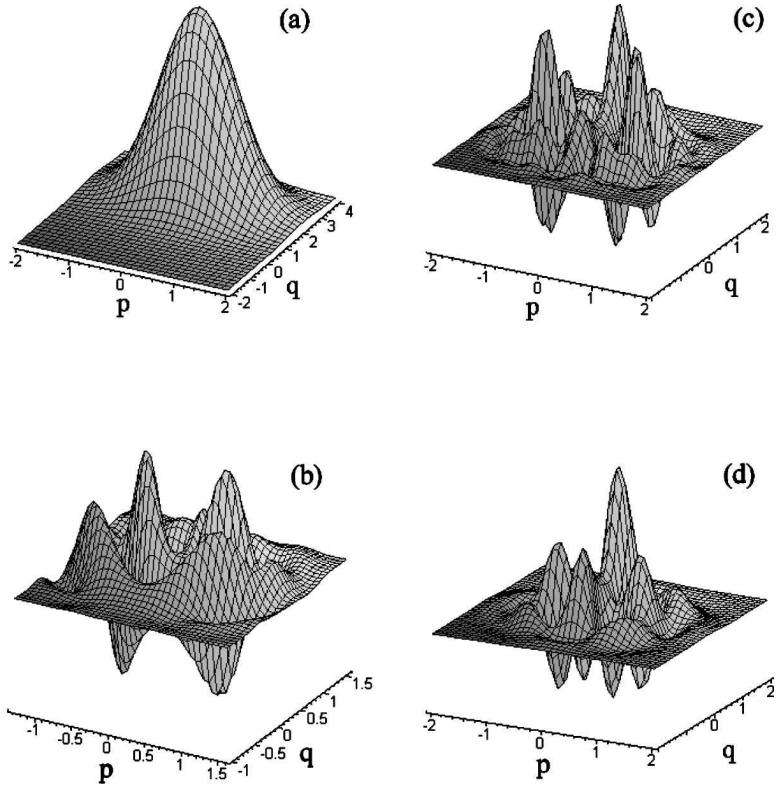

FIG. 3. Wigner distribution functions (a) for the coherent state associated with $\bar{n}_\alpha = 2.56$, (b) the cavity-field state after synthesizing the detection of the first atom, (c) the *sculptured* truncated phase state after synthesizing the detection of the second atom, and (d) the *desired* truncated phase state, respectively.

We are thankful for support from CNPq and FAPESP, Brazil, and we wish to thank R. J. Napolitano, M. Marchioli and P. Nussenzveig for helpful discussions.